\documentclass[12pt]{article}
\usepackage{amsfonts} 
\usepackage{amsmath}
\usepackage{amssymb}
\usepackage{amstext}
\usepackage{amsfonts}
\usepackage{graphicx,epsfig}
\usepackage{indentfirst}
\usepackage{hyperref}

\newtheorem{theorem}{Theorem}[section]
\newtheorem{lemma}[theorem]{Lemma}
\newtheorem{proposition}[theorem]{Proposition}

\newcommand{\U}{\mathrm{U}}
\newcommand{\GL}{\mathrm{GL}}
\newcommand{\M}{\mathrm{M}}
\newcommand{\rO}{\mathrm{O}}

\makeatletter
\def\ExtendSymbol#1#2#3#4#5{\ext@arrow 0099{\arrowfill@#1#2#3}{#4}{#5}}
\def\RightExtendSymbol#1#2#3#4#5{\ext@arrow 0359{\arrowfill@#1#2#3}{#4}{#5}}
\def\LeftExtendSymbol#1#2#3#4#5{\ext@arrow 6095{\arrowfill@#1#2#3}{#4}{#5}}
\makeatother

\begin{document}
\baselineskip 20pt

\title{Local unitary equivalence of quantum states and simultaneous
orthogonal equivalence}

\author {\small Naihuan Jing$^{1}$ \footnote{Email: jing@ncsu.edu} ,
Min Yang$^{1}$, Hui Zhao$^{2}$ \footnote{Corresponding author: zhaohui@bjut.edu.cn}\\[0.5cm]
\small \sl $^1$ Department of Mathematics, North Carolina State University, Raleigh, NC 27695, USA\\
\small \sl$^2$ College of Applied Science, Beijing University
of Technology, Beijing 100124, China}

\date{}
\maketitle

\begin{abstract}
The correspondence between local unitary equivalence of bipartite quantum states
and simultaneous orthogonal equivalence is thoroughly investigated and strengthened. It is proved that
local unitary equivalence can be studied through simultaneous similarity under
projective orthogonal transformations, and
four parametrization independent
algorithms are proposed to judge when
two density matrices on $\mathbb C^{d_1}\otimes \mathbb C^{d_2}$
are locally unitary equivalent in connection with
trace identities, Weierstrass pencils, Albert determinants and Smith normal forms.
\\


\end{abstract}



\section{ INTRODUCTION}

As one of the interesting and non-classic properties of quantum theory and information science \cite{HHHH, NC}, quantum entanglement has played an important role in
quantum computing \cite{RB},
quantum dense coding \cite{JZY}, quantum cryptography \cite{K} and quantum teleportation \cite{O}. It is necessary to determine and classify entanglement status of quantum states in quantum information theory. One step to solve this question is to determine the local unitary (LU) equivalence, as quantum entanglement
is invariant under LU transformation.

In \cite{Ma}, a set of 18 tensor invariants of local unitary equivalence is constructed for the 2-qubit mixed quantum state. Recently a refined set of 12 polynomial invariants \cite{J15} for generic $2$-qubits and 90 polynomial invariants
for generic $3$-qubits
have been found by using matrix elements of the Bloch representation. Nonlocal properties
 of multiqubits have been studied in \cite{LPS} long ago£¬ and a necessary and sufficient condition has been set up for the local unitary equivalence problem in multipartite pure qubits \cite{Kr1, Kr2}. In the case of bipartite qubits, a parametrization
dependent criterion for LU equivalence was given in \cite{Zhou}. While for multipartite quantum states, certain properties of LU equivalence are also considered in some special situations \cite{Li2, Liu}. In an indirect approach, generating sets
of local SL-equivalent classes are found for multipartite entanglements \cite{GW} and
abelian symmetry of the LU equivalence has been studied in \cite{J}.
It is also known that LU equivalence of density operators can be classified
using a finite set of polynomials \cite{YQS, MT} and spectrum-dependent bounds are given in
\cite{MOS}. Very recently a method to judge LU equivalence for multi-qubits \cite{Mat} was also
proposed and more generally SLOCC invariants
for multi-partite states are found \cite{JLLZF}. Despite all these developments, it remains a
challenging problem to effectively determine the LU equivalence by
an operational procedure using invariant polynomials. It is also noted that almost all previous methods
do not work for two particles with different dimensions.

In this paper, we strengthen the correspondence between the local unitary equivalence of bipartite quantum states
and simultaneous orthogonal equivalence of associated matrix triples and prove that
the local unitary equivalence can be transformed to the classical problem of
simultaneous projective orthogonality. We then introduce the concept of {\it quasi-LU equivalence} for bipartite states
using the latter matrix identities, while the quasi-LU equivalence becomes LU equivalence in the case of qubits.
Since the correspondence between simultaneous orthogonality and similarity
has been known in linear algebra \cite{J15b}, our new characterization simplifies
and emphasizes the connection with LU equivalence.
This enables us to give four algorithms to judge the local unitary equivalence of mixed bipartite quantum states
on any tensor product of two Hilbert spaces with dimensions not necessarily the same. In particular, we define a new canonical form
called {\it Smith normal form} for any bipartite quantum state, which provides a set of invariant polynomials for LU equivalent quantum states.
Moreover, our correspondence is completely general as it can treat LU equivalence for any two
particles over different dimensions.

One example is given to show how the algorithms uncovered in this work
are applied and the pros and cons of these algorithms are analyzed.
It is shown that through a finite procedure of checking trace identities
the problem of classifying LU equivalence can be completely settled for qubits. We then demonstrate that the set of
invariant polynomials given by the Smith normal form also provides effective necessary conditions for two qubit states being LU equivalent.

\section{LU Equivalence of Bipartite Quantum States}
Let $\rho$ be the density matrix of a bipartite state on $\mathcal{H}_{d_1}\otimes \mathcal{H}_{d_2}$, and let $\{ \lambda_i^{(k)}, 0\le i\le d_k^2-1, k=1,2\}$ be the Gell-Mann bases for each partite, then $\rho$ can be expressed in the following form:
\begin{align}\nonumber
\rho&=\frac{1}{d_1d_2} I_{d_1d_2}+\sum_{i=1}^{N_1} u_{i}\lambda_i^{(1)}\otimes \lambda_0^{(2)} +\sum_{j=1}^{N_2} v_{j} \lambda_0^{(1)}\otimes \lambda_j^{(2)}\\
&+\sum_{i=1}^{N_1}\sum_{j=1}^{N_2} w_{ij}\lambda_i^{(1)}\otimes \lambda_j^{(2)},\quad N_k=d_k^2-1, k=1,2
\end{align}
where
$
u_i=\langle \rho , \lambda_i^{(1)}\otimes \lambda_0^{(2)}\rangle=tr\rho (\lambda_i^{(1)}\otimes\lambda_0^{(2)})$,
$v_j=\langle \rho , \lambda_0^{(1)}\otimes \lambda_j^{(2)}\rangle=tr\rho (\lambda_0^{(1)}\otimes\lambda_j^{(2)})$,
$w_{ij}=\langle \rho, \lambda_i^{(1)}\otimes \lambda_j^{(2)}\rangle=tr\rho (\lambda_i^{(1)}\otimes\lambda_j^{(2)})$.
We associate three matrices for $\rho$:
\begin{align}
u(\rho)=[u_1, u_2,\cdots, u_{N_1}]^t, \quad v(\rho)=[v_1,v_2,\cdots, v_{N_2}]^t,\quad
W(\rho)=[w_{ij}]_{N_1\times N_2}.
\end{align}
and call them a {\it matrix representation} of the density matrix $\rho$.
For convenience we denote $UM U^\dagger$ by $M^U$, where $M\in\M(n), U\in\U(n)$.

Suppose $\rho'=\rho^{U_1\otimes U_2}$ is another mixed bipartite state on $\mathcal{H}_{d_1}\otimes \mathcal{H}_{d_2}$
for two unitary matrices $U_i\in\U(d_i)$. Therefore we can write that
$(\lambda_i^{(1)})^{U_1}=\sum_{j=1}^{N_1} a_{ij}\lambda_j^{(1)}$, $(\lambda_i^{(2)})^{U_2}=\sum_{j=1}^{N_2} b_{ij}\lambda_j^{(2)}$
for two complex matrices $A$ and $B$, and one sees that
\begin{align}
\sum_{i=1}^{N_1} u_i(\lambda_i^{(1)})^{U_1}&\otimes \lambda_0^{(2)}=\sum_{i=1}^{N_1}\sum_{j=1}^{N_2} u_ia_{ij}\lambda_j^{(1)}\otimes \lambda_0^{(2)}
=\sum_{i=1}^N(\sum_{j=1}^{N_2} u_ja_{ji})\lambda_i^{(1)}\otimes \lambda_0^{(2)},
\end{align}
i.e.\; $u(\rho^{U_1\otimes U_2})=A^t u(\rho)$. Similarly $v(\rho^{U_1\otimes U_2})=B^t v(\rho)$, and $ W(\rho^{U_1\otimes U_2})=A^t W(\rho)B$.

\begin{lemma}\label{l:tech} Let $\rho$ and $\rho'$ be two locally unitary equivalent density matrices, then there exist two
real orthogonal matrices $A\in \rO(N_1)$ and $B\in \rO(N_2)$ such that $u(\rho')=A^tu(\rho)$, $v(\rho')=B^tv(\rho)$, and $W(\rho')=A^tW(\rho) B.$
\end{lemma}
\noindent{\bf Proof}. Let $\{\lambda_i\}$ be an orthonormal hermitian basis in $End(V)$ under the trace form,
and let $U$ be a unitary matrix $\in End(V)$.
Write $\lambda_i^U=U\lambda_iU^{\dagger}=\sum_{ij}m_{ij}\lambda_j$.
As $(\lambda_i^U)^{\dagger}=U\lambda_i^{\dagger}U^{\dagger}=U\lambda_iU^{\dagger}$, the coefficients $m_{ij}$ are real numbers.
The orthogonality of $\{\lambda_i^U\}$ is an easy consequence of the following computation:
$$tr(\lambda_i^U\lambda_j^U)=tr(U\lambda_i\lambda_j U^\dagger)=tr(\lambda_i\lambda_jU^\dagger U)=tr(\lambda_i\lambda_j)=\delta_{ij}. $$
By a general result of linear algebra, any two orthonormal bases are transformed by an orthogonal matrix, therefore $MM^T=M^TM=I$.
The matrix equations have already been verified above. $\Box$

Two bipartite density matrices $\rho_1$ and $\rho_2$ over $\mathbb C^{d_1}\otimes \mathbb C^{d_2}$ are then called {\it quasi local unitary equivalent} if there exist two orthogonal matrices $O_1, O_2$, $O_i\in \mathrm{O}(d_i^2-1)$ such that
\begin{align}\label{quasiLU}
u(\rho_2)=O_1u(\rho_1), \quad v(\rho_2)=O_2v(\rho_1), \quad W(\rho_2)=O_1W(\rho_1)O_2^t.
\end{align}

By Lemma \ref{l:tech} two LU equivalent bipartite mixed states are quasi-LU equivalent. In the case of two qubits, it is well-known that
 quasi-LU equivalence is also a sufficient condition for LU equivalence (see for example, \cite{J15}).

\section{Criteria of Simultaneous Orthogonal Equivalence}

Suppose $\{W_i, u_i, v_i\}$ is a matrix representation of the density matrix $\rho_i$ on
$\mathbb C^{d_1}\otimes \mathbb C^{d_2}$, where
$W_i$ is an $m\times n$ matrix and $u_i, v_i$ are column vectors of dimension $m$ and $n$ respectively
(here $m=d_1^2-1, n=d_2^2-1$).

By the remark after Lemma \ref{l:tech}, two qubits
$\rho_1$ and $\rho_2$ are LU equivalent if and only if there are orthogonal matrices $O_i$ such that
$O_1W_1O_2^t=W_2$, $O_1u_1=u_2$, and $O_2v_1=v_2$, it follows that
the set $\{W_1^tW_1, v_1u_1^t\}$ is simultaneously orthogonally equivalent to
$\{W_2^tW_2, v_2u_2^t\}$. However, the converse direction is not true in general.

We first give a simplified correspondence between the quasi-LU equivalence
and the projective orthogonal equivalence of two real matrices. In particular, it implies that
under a norm condition if $\{W_1^tW_1, v_1u_1^t\}$ is simultaneously orthogonal equivalent to $\{W_2^tW_2, v_2u_2^t\}$,
then the two mixed states $\rho_1$ and $\rho_2$ are quasi-LU equivalent.

\begin{theorem}
{\it (Correspondence between quasi-LU and simultaneous orthogonal equivalence)}.
Let $W_i\in\mathbb R_{m\times n}$, $u\in \mathbb R_m$ and $v\in \mathbb R_n$.
There exist orthogonal matrices $O_1\in \rO(m)$ and $O_2\in \rO(n)$ such that
$O_1W_1O_2^t=W_2, O_1u_1=u_2, O_2v_1=v_2$
 if and only if $\{W_1, u_1v_1^t\}$ is simultaneously orthogonal equivalent to $\{W_2, u_2v_2^t\}$
 and $|u_1|=|u_2|$ or $|v_1|=|v_2|$.
\end{theorem}
\noindent{\bf Proof}. The necessity has already been checked.
Suppose there exist two orthogonal matrices $O_i$ such that
$
O_1W_1O_2^t=W_2, \quad O_1u_1v_1^tO_2^t=u_2v_2^t.
$
Without loss of generality we can assume that both $u_2, v_2\neq 0$. Note that
$
u_2^tO_1u_1v_1^tO_2^tv_2=u_2^tu_2v_2^tv_2\neq 0,
$
then $\alpha=\frac{v_2^tv_2}{v_1^tO_2^tv_2}=\frac{u_2^tO_1u_1}{u_2^tu_2}\neq 0$, which implies that
$
O_1u_1=\alpha u_2, \quad O_2v_1=\alpha^{-1} v_2.
$
Let $\overline{O_1}=\alpha^{-1}O_1, \overline{O_2}=\alpha O_2$. As $|u_1|=|u_2|$ or $|v_1|=|v_2|$, we see that
$\alpha=\pm 1$. Then
$
\overline{O}_1W_1\overline{O}_2^t=W_2, \quad \overline{O}_1u_1=u_2, \overline{O}_2v_1=v_2,
$
where $\overline{O}_i\in \rO(n)$. 
$\Box$.


Through this correspondence, we have transformed the LU problem to that of simultaneous orthogonal equivalence
between two pairs of matrices plus the norm condition.
To solve this problem, we first look at several algorithms
to judge when two sets of real matrices simultaneously orthogonal
similar, and then reduce the problem of simultaneous orthogonal equivalence to that
of simultaneous (orthogonal) similarity, which is one of the classical problems in linear algebra.

Recall that two square matrices $A$ and $B$ are {\it similar} if there exists an orthogonal matrix
$O$ such that $A=OBO^t$. The fundamental {\it Specht's criterion} \cite{Specht} says that
a square matrix $A$ is similar to $B$ if and only if
\begin{equation}
trw(A, A^{\dagger})=trw(B, B^{\dagger})
\end{equation}
for any word $w(x, y)=x^{m_1}y^{n_1}\cdots x^{m_k}y^{n_k}$, where $m_i, n_i\in\mathbb Z_+$ and $k\in \mathbb N$.
Specht's criterion has been generalized to two sets of normal matrices \cite{Alpin, J15b}, where
the trace identities are for all words in the alphabet of the matrix set and the transpose.
We can give our second result to judge when two density matrices are quasi-LU equivalent.

\begin{theorem}
{\it (Simultaneous orthogonal equivalence)}. Let $\rho_i$ be two bipartite
density matrices over the same Hilbert space and suppose $\{W_i, u_i, v_i\}$ are the associated matrix triples. Let
$\{A_1, A_2, A_3\}=\{W_1W_1^t,
W_1v_1u_1^t, u_1u_1^t\}$, and $\{B_1, B_2, B_3\}=\{W_2W_2^t, W_2v_2u_2^t, u_2u_2^t\}.$
Then $\rho_1$ and $\rho_2$ are quasi-LU equivalent if and only if the trace identities hold:
\begin{equation}\label{e:tr}
tr(A_{i_1}A^t_{j_1}\cdots A_{i_k}A_{j_k}^t)=tr(B_{i_1}B_{j_1}^t\cdots B_{i_k}B_{j_k}^t)
\end{equation}
for any compositions $i_1, \cdots, i_k$ and $j_1, \cdots, k_k$ of $\{1, 2, 3\}$ such that
$1\leq i_1\leq j_1\leq 3, \cdots, 1\leq i_k\leq j_k\leq 3.
$
Moreover, Eq. (\ref{e:tr}) are sufficient conditions for LU equivalence in the case of two qubits.
\end{theorem}
\noindent{\bf Proof}. First of all, from our previous discussion it follows that
if two density matrices $\rho_1$ and $\rho_2$ are quasi-LU equivalent, then
$\{W_i, u_iv_i^t\}$ are simultaneous orthogonal equivalent and $|u_1|=|u_2|$.
Subsequently the sets $\{W_iW_i^t, W_iv_iu_i^t, u_iu_i^t\}$ are simultaneously orthogonal similar. According to \cite[Th. 3.3]{J15b} two sets of rectangular matrices
$\{A_1, \cdots, A_l\}$ and $\{B_1, \cdots, B_l\}$ of the same size are orthogonally equivalent
if and only if the equations in \eqref{e:tr} hold for any compositions $i_1, \cdots, i_k$
of the integers $\{1, \cdots, k\}$ such that $1\leq i_t\leq j_t\leq k$, $t=1, \cdots, k$.
So the theorem is proved.
$\Box$

There is a simpler necessary condition arising from the
connection with Jordan algebras \cite{Al}, which can be proved directly.

\begin{theorem} {\it (Albert's criterion)} Suppose that $(W, u, v)$ is a matrix representation
of the density matrix $\rho$, then
\begin{equation}\det(xI-x_1WW^t-x_2uu^t-x_3Wvu^t)\end{equation}  
is an invariant polynomial in the $x_i$ under the LU equivalence. This partly generalizes Makhlin's invariants.
\end{theorem}

We remark that Albert's criterion is a natural generalization of the characteristic polynomial
of a square matrix.
It is also easy to see that the generalized characteristic polynomial given by Albert contains several
invariants considered by Makhlin \cite{Ma}.

  Gerasimova, Horn and Sergeichuk \cite{GHS} gave an algorithm to judge
simultaneous orthogonal similarity using block matrices to reduce the problem to Specht's criterion.
We reformulate it as follows.

\begin{theorem} {\it (GHS algorithm)}. Suppose $m\leq n$, and consider the nilpotent matrices
\begin{equation}\begin{bmatrix}
0 & I_m & u_1u_1^t & W_1W_1^t\\
 & 0 & I_m & W_1v_1u_1^t\\
 &  & 0 & I_m\\
 &  &  & 0
 \end{bmatrix},
\quad
\begin{bmatrix}
0 & I_m & u_2u_2^t & W_2W_2^t\\
 & 0 & I_m & W_2v_2u_2^t\\
 &  & 0 & I_m\\
 &  &  & 0
 \end{bmatrix}
\end{equation}
Then these two $4m\times 4m$ matrices are orthogonal similar if and only if
the set $\{W_1W_1^t, u_1u_1^t, W_1v^tu_1\}$ is simultaneous orthogonal equivalent to the set
$\{W_2W_2^t, u_2u_2^t, W_2v^tu_2\}$ or $\rho_1$ and $\rho_2$ are quasi-LU equivalent.
\end{theorem}
\noindent{\bf Proof}. The criterion is directly checked by working out the matrix product and see that
the equations of entries imply the simultaneous orthogonality.
$\Box$

The GHS algorithm transforms the LU problem into that of orthogonality similarity between two
block matrices.

\section{Smith Normal Forms of Kronecker Pencils} 
We now introduce an effective criterion for simultaneous similarity of the
triple matrices.
Let $\rho$ be a density matrix on  $H_{d_1}\otimes H_{d_2}$ associated with $(W(\rho),u(\rho),v(\rho))$, we consider the auxiliary $\lambda$-matrix $\lambda W(\rho)+u(\rho)v(\rho)^t$ known as {\it the Kronecker pencil} \cite{Ga}.
As an element of the ring $\mathbb C[\lambda]$ of matrix polynomials in $\lambda$, the $\lambda$-matrix $\lambda W(\rho)+u(\rho)v(\rho)^t$ is equivalent to the Smith normal form \cite{Ga, Ja} under elementary row/column operations.
The Smith normal form is defined by the property that it is a diagonal matrix over $\mathbb C[\lambda]$
and each non-zero
main diagonal entry divides its next diagonal entry. It is normalized such that the diagonal entries
$d_i(\lambda)$ are monic polynomials, i.e.
\begin{equation}S(\lambda)=\left[\begin{matrix}
d_1(\lambda) & 0 & \cdots & 0 & 0 &\cdots\\
0 & d_2(\lambda) & \cdots & 0 & 0 &\cdots\\
\cdots & \cdots & \cdots & \cdots &\cdots &\cdots\\
0& 0 & \cdots & d_m(\lambda) & 0 &\cdots\\
0 & 0 & \cdots & 0 & 0 &\cdots\\
\cdots & \cdots & \cdots & \cdots & \cdots & \cdots
\end{matrix}\right]_{N\times M},\end{equation}
where $d_i(\lambda) \in \mathbb{C}[\lambda]$ such that $d_i(\lambda)|d_{i+1}(\lambda)$, $\forall 1\le i\le m$ and $m\le N.$
As $d_i(\lambda)$ are successively given by the principal minors of $\lambda$-matrix $\lambda W(\rho)+u(\rho)v(\rho)^t$, the Smith normal form is uniquely determined and invariant under
elementary operations. Furthermore it is well-known \cite{Ja} that there are
$P(\lambda)\in \GL_{N_1}(\mathbb{C}[\lambda])$, $Q(\lambda)\in \GL_{N_2}(\mathbb{C}[\lambda])$ such
 that $P(\lambda)(\lambda W(\rho)+u(\rho)v(\rho)^t)Q(\lambda)=S(\lambda)$.

Let $\rho$ be a density matrix for bipartite system over $H_{d_1}\otimes H_{d_2}$ with the matrix representation
$(W(\rho),u(\rho),v(\rho))$, we will simply call the Smith normal form of $\lambda W(\rho)+u(\rho)v(\rho)^t$
as the {\it Smith normal form} of the triple $(W(\rho),u(\rho),v(\rho))$.
We can now state our fourth criterion.

\begin{theorem}\label{Smith}
({\it Smith Normal Form}). For any two bipartite quantum states $\rho, \rho'$ associated with $(W(\rho), u(\rho), v(\rho))$ and $(W(\rho'), u(\rho'), v(\rho'))$ respectively. If $\rho$ is local unitary equivalent to $\rho'$, then the Smith
normal forms of the triple systems $(W(\rho), u(\rho), v(\rho))$ and $(W(\rho'), u(\rho'), v(\rho'))$ are the same.
\end{theorem}

We remark that the normal form of a $\lambda$-matrix was introduced years ago by Weierstrass for
regular cases, by Kronecker for singular cases \cite{Ga}, and in general by Smith \cite{Ja}.
It should not be confused with the much younger term of the canonical form given by the Schmidt decomposition in quantum computation.

\begin{proposition}\label{pencil}
For any matrices $X, X', Y$ and $Y'\in \M_N(\mathbb{C})$, there exists $U_1, U_2\in \U(N)$ (or $\rO(N)$), such that $U_1XU_2^\dagger=X', U_1YU_2^\dagger=Y'$ if and only if there exists $U_1, U_2\in \U(N)$ (or $\rO(N)$), such that $U_1(X+\lambda Y)U_2^\dagger=(X'+\lambda Y').$
\end{proposition}
\noindent{\bf Proof}. This sufficient direction can be easily seen as follows.
Suppose there exist $U_1, U_2\in \U(N)$ such that $U_1(X+\lambda Y)U_2^\dagger=(X'+\lambda Y')$.
Let $\lambda=0, 1$, we obtain that $U_1XU_2^\dagger=X'$ and $U_1(X+Y)U_2^\dagger=X'+Y'$. Taking
difference, the other equation is also obtained. $\Box$

\noindent{\bf Proof of Smith normal form}.
We know that if $\rho$ is equivalent to $\rho'$ there exist two orthogonal matrices $U_i$ in $\mathrm{O}(N_i)$ such that $u(\rho')=U_1u(\rho)$, $v(\rho')=U_2 v(\rho)$, $W(\rho')=U_1W(\rho) U_2^t$. It follows that $U_1u(\rho)v(\rho)^\dagger U_2^\dagger=u(\rho')v(\rho')^\dagger $.
Subsequently
\begin{equation}U_1(\lambda W(\rho)+ u(\rho)v(\rho)^\dagger)U_2^\dagger=\lambda W(\rho')+ u(\rho')v(\rho')^\dagger, \end{equation}
thus they have the same normal form. $\Box$

\medskip

 Suppose $\lambda W_1(\rho)+ u_1(\rho)v_1(\rho)^t$ and $\lambda W_2(\rho)+ u_2(\rho)v_2(\rho)^t$
 have the same Smith normal form. Then there are invertible matrices $P(\lambda)$ and $Q(\lambda)$ such that
\begin{equation}
P(\lambda)(\lambda W_1(\rho)+ u_1(\rho)v_1(\rho)^t)Q(\lambda)=\lambda W_2(\rho)+ u_2(\rho)v_2(\rho)^t.
\end{equation}
Since $P(\lambda)$ and $Q(\lambda)$ are
obtained by Gauss elimination, $P(\lambda)$ and $Q(\lambda)$ are polynomial functions of $\lambda$
with non-zero constants. In fact the constants must be invertible matrices. Therefore one obtains
that
\begin{equation}
P(\lambda W_1(\rho)+ u_1(\rho)v_1(\rho)^t)Q=\lambda W_2(\rho)+ u_2(\rho)v_2(\rho)^t.
\end{equation}
for two invertible matrices $P, Q$. i.e. They are strictly equivalent in the sense of Gantmacher \cite{Ga}.

{\it Example}. Consider the following quantum state $\rho$ in $\mathcal{H}_{2}\otimes \mathcal{H}_{3}$.
\begin{eqnarray}
  \rho &=& \frac{1}{6}I_{6}+ \frac{1-p}{3}\lambda_{1}^{(1)}\otimes\lambda_{0}^{(2)}-\frac{1-p}{2}\lambda_{0}^{(1)}\otimes\lambda_{3}^{(2)}
  +\frac{1}{2\sqrt{3}}\lambda_{0}^{(1)}\otimes\lambda_{8}^{(2)} \nonumber\\[3mm]
  &&+\frac{2p-1}{2}\lambda_{1}^{(1)}\otimes\lambda_{3}^{(2)}+\frac{1-p}{2\sqrt{3}}\lambda_{1}^{(1)}\otimes\lambda_{8}^{(2)}
  +\frac{p}{2}\lambda_{2}^{(1)}\otimes\lambda_{1}^{(2)}+\frac{p}{2}\lambda_{3}^{(1)}\otimes\lambda_{2}^{(2)},
\end{eqnarray}  where $p\in [0,1]$ and $\lambda_0^{(1)}=I_2/\sqrt{2}$, $\lambda_1^{(1)}=(|0\rangle\langle 0|-|1\rangle\langle 1|)/\sqrt{2}, \lambda_2^{(1)}=(|0\rangle\langle 1|+|1\rangle\langle 0|)/\sqrt{2}$, $\lambda_3^{(1)}=(i|0\rangle\langle 1|-i|1\rangle\langle 0|)/\sqrt{2}$,
$\lambda_0^{(2)}=I_3/\sqrt{2}$, $\lambda_{1}^{(2)} = \frac{1}{\sqrt{2}}(|0\rangle\langle1|+|1\rangle\langle0|)$,
 $\lambda_{2}^{(2)} = -\frac{i}{\sqrt{2}}(|0\rangle\langle1|-|1\rangle\langle0|)$,
 $\lambda_{3}^{(2)} = \frac{1}{\sqrt{2}}(|0\rangle\langle0|-|1\rangle\langle1|)$,
 $\lambda_{8}^{(2)} = \frac{1}{\sqrt{6}}(|0\rangle\langle0|+|1\rangle\langle1|-2|2\rangle\langle2|)$.
Three matrices for $ \rho$ are
$\mu(\rho) = (\frac{1-p}{3},0,0)^{T}$,
$\nu(\rho) = (0,0,-\frac{1-p}{2},0,0,0,0,\frac{1}{2\sqrt{3}})^{T}$, and
$$W(\rho) = \left(
                           \begin{array}{cccccccc}
                             0 & 0 & \frac{2p-1}{2} & 0 &0 & 0 & 0 & \frac{1-p}{2\sqrt{3}} \\
                             \frac{p}{2} & 0& 0& 0& 0& 0& 0& 0 \\
                              0 &\frac{p}{2} & 0 & 0& 0 & 0& 0 & 0 \\
                           \end{array}
                         \right).
                         $$
Suppose $ \rho$ is local unitary equivalent to $ \rho'$ under 
$$U_{1}\otimes U_{2}=\left(
\begin{array}{cc}
\frac{1}{\sqrt{2}} & \frac{1}{\sqrt{2}}i \\
 -\frac{1}{\sqrt{2}}i & -\frac{1}{\sqrt{2}} \\
 \end{array}
 \right)\otimes\left(
 \begin{array}{ccc}
  0 & 1 & 0 \\
  1 & 0& 0 \\
  0 & 0& 1 \\
  \end{array}
   \right),$$
three associated matrices for $\rho'$ are
$\mu(\rho') = (0,0,\frac{1-p}{3})^{T}$,
$\nu(\rho') = (0,0,\frac{1-p}{2},0,0,0,0,\frac{1}{2\sqrt{3}})^{T}$, and
$$W(\rho') = \left(
                           \begin{array}{cccccccc}
                            0 &-\frac{p}{2} & 0& 0 & 0& 0 & 0& 0  \\
                             -\frac{p}{2} & 0 & 0 & 0 & 0& 0 & 0& 0 \\
                              0 & 0 & -\frac{2p-1}{2}& 0 & 0& 0 & 0&\frac{1-p}{2\sqrt{3}}  \\
                           \end{array}
                         \right).$$
Then there exist orthogonal matrices
\begin{eqnarray}A=\left(
                      \begin{array}{ccc}
                        0 & 0 & -1 \\
                        0 & -1 & 0 \\
                        1 & 0 & 0 \\
                      \end{array}
                    \right),\ \
B=\left(
                      \begin{array}{cccccccc}
                         1 & 0 & 0& 0 & 0& 0 & 0& 0 \\
                        0 & 1 & 0& 0 & 0& 0 & 0& 0  \\
                       0 & 0 & -1 & 0 & 0& 0 & 0& 0 \\
                       0 & 0 & 0& 1 & 0& 0 & 0& 0 \\
                       0 & 0 & 0& 0 & 1& 0 & 0& 0 \\
                       0 & 0 & 0& 0 & 0& 1 & 0& 0 \\
                       0& 0 & 0& 0 & 0& 0 & 1& 0 \\
                       0 & 0 & 0& 0 & 0& 0 & 0& 1 \\
                      \end{array}
                    \right)\end{eqnarray}
such that
$\mu(\rho')=A^{T}\mu(\rho)$, $\nu(\rho') =B^{T}\nu(\rho)$, $W(\rho') =A^{T} W(\rho)B$.
And $\rho$ and $\rho'$ have the same Smith normal forms
$$\left(
         \begin{array}{cccccccc}
          1 & 0 & 0& 0 & 0& 0 & 0& 0 \\
           0 & \frac{p}{2}\lambda & 0& 0& 0& 0& 0& 0 \\
            0 & 0 & \frac{p}{2}\lambda & 0& 0& 0& 0& 0  \\
         \end{array}
       \right).$$
Moreover the polynomial
\begin{eqnarray}
&&det(xI-x_1W(\rho)W(\rho)^t-x_2u(\rho)u(\rho)^t-x_3W(\rho)v(\rho)u(\rho)^t) \nonumber\\
&=&det(xI-x_1W(\rho')W(\rho')^t-x_2u(\rho')u(\rho')^t-x_3W(\rho')v(\rho')u(\rho')^t) \nonumber\\
&=&(x-\frac{p^2}{4}x_1)^2(x-\frac{3(2p-1)^2+(1-p)^2}{12}x_1-\frac{(1-p)^2}{9}x_2-\frac{(2-3p)(1-p)^2}{18}x_3)
\end{eqnarray}
is an invariant polynomial of the LU equivalence.
It can be directly checked that
$$\begin{bmatrix}
0 & I_3 & u(\rho_1)u(\rho_1)^T & W(\rho_1)W(\rho_1)^T\\
 & 0 & I_3 & W(\rho_1)v(\rho_1)u(\rho_1)^T\\
 &  & 0 & I_3\\
 &  &  & 0
 \end{bmatrix}$$
$$ =I_{12}^T
\begin{bmatrix}
0 & I_3 & u(\rho_{1}')u(\rho_{1}')^T & W(\rho_{1}')W(\rho_{1}')^T\\
 & 0 & I_3 & W(\rho_{1}')v(\rho_{1}')u(\rho_{1}')^T\\
 &  & 0 & I_3\\
 &  &  & 0
 \end{bmatrix}I_{12},$$
according to the GHS algorithm, two $12\times 12$ nilpotent matrices are orthogonal similar
under a matrix. We find that this matrix is given by 
$$\begin{bmatrix}P & 0 & 0 & 0\\
 & P & 0 & 0\\
 &  & P & 0\\
 &  &  & P
\end{bmatrix}, 
\quad P=\begin{bmatrix}0 & 0 & 1\\
 0 & 1 & 0\\
 1& 0 & 0
\end{bmatrix}. $$
Then $\{WW^t, uu^t, Wvu^t\}$ is simultaneous orthogonal similar
to $\{W'W^{'T}, u'u'^{'T}, W'v'u^{'T}\}$.


\bigskip


\section*{ACKNOWLEDGMENTS}
This work is partially supported by Simons Foundation grant No. 198129, National Natural Science Foundation of China
grant Nos. 11101017, 11281137, 11531004 and
an award from China Scholarship Council.

\end{document}